\PassOptionsToPackage{unicode}{hyperref}
\PassOptionsToPackage{hyphens}{url}
\PassOptionsToPackage{dvipsnames,svgnames,x11names}{xcolor}
\documentclass[
  12pt,
]{interact}

\usepackage{amsmath,amssymb}
\usepackage{setspace}
\usepackage{iftex}
\ifPDFTeX
  \usepackage[T1]{fontenc}
  \usepackage[utf8]{inputenc}
  \usepackage{textcomp} 
\else 
  \usepackage{unicode-math}
  \defaultfontfeatures{Scale=MatchLowercase}
  \defaultfontfeatures[\rmfamily]{Ligatures=TeX,Scale=1}
\fi
\usepackage{lmodern}
\ifPDFTeX\else  
\fi
\IfFileExists{upquote.sty}{\usepackage{upquote}}{}
\IfFileExists{microtype.sty}{
  \usepackage[]{microtype}
  \UseMicrotypeSet[protrusion]{basicmath} 
}{}
\makeatletter
\@ifundefined{KOMAClassName}{
  \IfFileExists{parskip.sty}{%
    \usepackage{parskip}
  }{
    \setlength{\parindent}{0pt}
    \setlength{\parskip}{6pt plus 2pt minus 1pt}}
}{
  \KOMAoptions{parskip=half}}
\makeatother
\usepackage{xcolor}
\ifx\paragraph\undefined\else
  \let\oldparagraph\paragraph
  \renewcommand{\paragraph}[1]{\oldparagraph{#1}\mbox{}}
\fi
\ifx\subparagraph\undefined\else
  \let\oldsubparagraph\subparagraph
  \renewcommand{\subparagraph}[1]{\oldsubparagraph{#1}\mbox{}}
\fi

\usepackage{color}
\usepackage{fancyvrb}

\DefineVerbatimEnvironment{Highlighting}{Verbatim}{commandchars=\\\{\}}
\usepackage{framed}
\definecolor{shadecolor}{RGB}{241,243,245}
\newenvironment{Shaded}{\begin{snugshade}}{\end{snugshade}}

\newcommand{\AttributeTok}[1]{\textcolor[rgb]{0.40,0.45,0.13}{#1}}

\newcommand{\ControlFlowTok}[1]{\textcolor[rgb]{0.00,0.23,0.31}{#1}}

\newcommand{\DecValTok}[1]{\textcolor[rgb]{0.68,0.00,0.00}{#1}}

\newcommand{\FunctionTok}[1]{\textcolor[rgb]{0.28,0.35,0.67}{#1}}

\newcommand{\NormalTok}[1]{\textcolor[rgb]{0.00,0.23,0.31}{#1}}

\newcommand{\OtherTok}[1]{\textcolor[rgb]{0.00,0.23,0.31}{#1}}

\newcommand{\SpecialCharTok}[1]{\textcolor[rgb]{0.37,0.37,0.37}{#1}}

\newcommand{\StringTok}[1]{\textcolor[rgb]{0.13,0.47,0.30}{#1}}

\providecommand{\tightlist}{%
  \setlength{\itemsep}{0pt}\setlength{\parskip}{0pt}}\usepackage{longtable,booktabs,array}
\usepackage{calc} 
\usepackage{etoolbox}
\makeatletter
\patchcmd\longtable{\par}{\if@noskipsec\mbox{}\fi\par}{}{}
\makeatother
\IfFileExists{footnotehyper.sty}{\usepackage{footnotehyper}}{\usepackage{footnote}}
\makesavenoteenv{longtable}
\usepackage{graphicx}
\makeatletter
\def\maxwidth{\ifdim\Gin@nat@width>\linewidth\linewidth\else\Gin@nat@width\fi}
\def\maxheight{\ifdim\Gin@nat@height>\textheight\textheight\else\Gin@nat@height\fi}
\makeatother
\setkeys{Gin}{width=\maxwidth,height=\maxheight,keepaspectratio}
\makeatletter
\def\fps@figure{htbp}
\makeatother
\NewDocumentCommand\citeproctext{}{}

\makeatletter
 \let\@cite@ofmt\@firstofone
 \def\@biblabel#1{}
 \def\@cite#1#2{{#1\if@tempswa , #2\fi}}
\makeatother
\newlength{\cslhangindent}
\newlength{\csllabelwidth}
\newenvironment{CSLReferences}[2] 
 {\begin{list}{}{%
  \setlength{\itemindent}{0pt}
  \setlength{\leftmargin}{0pt}
  \setlength{\parsep}{0pt}
  \ifodd #1
   \setlength{\leftmargin}{\cslhangindent}
   \setlength{\itemindent}{-1\cslhangindent}
  \fi
  \setlength{\itemsep}{#2\baselineskip}}}
 {\end{list}}
\usepackage{calc}

\usepackage{booktabs}
\usepackage{longtable}
\usepackage{array}
\usepackage{multirow}
\usepackage{wrapfig}
\usepackage{float}
\usepackage{colortbl}
\usepackage{pdflscape}
\usepackage{tabu}
\usepackage{threeparttable}
\usepackage{threeparttablex}
\usepackage[normalem]{ulem}
\usepackage{makecell}
\usepackage{xcolor}
\usepackage{orcidlink}
\makeatletter
\@ifpackageloaded{caption}{}{\usepackage{caption}}
\AtBeginDocument{%
\ifdefined\contentsname
  \renewcommand*\contentsname{Table of contents}
\else
  \newcommand\contentsname{Table of contents}
\fi
\ifdefined\listfigurename
  \renewcommand*\listfigurename{List of Figures}
\else
  \newcommand\listfigurename{List of Figures}
\fi
\ifdefined\listtablename
  \renewcommand*\listtablename{List of Tables}
\else
  \newcommand\listtablename{List of Tables}
\fi
\ifdefined\figurename
  \renewcommand*\figurename{Figure}
\else
  \newcommand\figurename{Figure}
\fi
\ifdefined\tablename
  \renewcommand*\tablename{Table}
\else
  \newcommand\tablename{Table}
\fi
}
\@ifpackageloaded{float}{}{\usepackage{float}}
\floatstyle{ruled}
\@ifundefined{c@chapter}{\newfloat{codelisting}{h}{lop}}{\newfloat{codelisting}{h}{lop}[chapter]}
\floatname{codelisting}{Listing}

\makeatother
\makeatletter
\@ifpackageloaded{caption}{}{\usepackage{caption}}
\@ifpackageloaded{subcaption}{}{\usepackage{subcaption}}
\makeatother
\ifLuaTeX
  \usepackage{selnolig}  
\fi
\usepackage{bookmark}

\IfFileExists{xurl.sty}{\usepackage{xurl}}{} 
\hypersetup{
  pdftitle={A Tidy Framework and Infrastructure to Systematically Assemble Spatio-temporal Indexes from Multivariate Data},
  pdfauthor={H. Sherry Zhang; Dianne Cook; Ursula Laa; Nicolas Langrené; Patricia Menéndez},
  pdfkeywords={indexes, data pipeline, software
design, uncertainty, decision-making},
  colorlinks=true,
  linkcolor={blue},
  filecolor={Maroon},
  citecolor={Blue},
  urlcolor={Blue},
  pdfcreator={LaTeX via pandoc}}

\title{A Tidy Framework and Infrastructure to Systematically Assemble
Spatio-temporal Indexes from Multivariate Data}
\author{H. Sherry
Zhang$\textsuperscript{1,2}$~\orcidlink{0000-0002-7122-1463}, Dianne
Cook$\textsuperscript{2}$~\orcidlink{0000-0002-3813-7155}, Ursula
Laa$\textsuperscript{3}$~\orcidlink{0000-0002-0249-6439}, Nicolas
Langrené$\textsuperscript{4}$~\orcidlink{0000-0001-7601-4618}, Patricia
Menéndez$\textsuperscript{5}$~\orcidlink{0000-0003-0701-6315}}

\thanks{CONTACT: H. Sherry
Zhang. Email: \href{mailto:huize.zhang@austin.utexas.edu}{\nolinkurl{huize.zhang@austin.utexas.edu}}. }
\begin{document}
\captionsetup{labelsep=space}
\maketitle
\textsuperscript{1} Department of Statistics and Data
Sciences, University of Texas at Austin, Austin,
Texas, USA\\ \textsuperscript{2} Department of Econometrics and Business
Statistics, Monash University, Melbourne,
Victoria, Australia\\ \textsuperscript{3} Institute of
Statistics, University of Natural Resources and Life
Sciences, Vienna, Austria\\ \textsuperscript{4} Guangdong Provincial Key
Laboratory of Interdisciplinary Research and Application for Data
Science, BNU-HKBU United International College, Zhuhai,
Guangdong, China\\ \textsuperscript{5} School of Mathematics and
Statistics, University of Melbourne, Melbourne, Victoria, Australia
\begin{abstract}
Indexes are useful for summarizing multivariate information into single
metrics for monitoring, communicating, and decision-making. While most
work has focused on defining new indexes for specific purposes, more
attention needs to be directed towards making it possible to understand
index behavior in different data conditions, and to determine how their
structure affects their values and the variability therein. Here we
discuss a modular data pipeline recommendation to assemble indexes. It
is universally applicable to index computation and allows investigation
of index behavior as part of the development procedure. One can compute
indexes with different parameter choices, adjust steps in the index
definition by adding, removing, and swapping them to experiment with
various index designs, calculate uncertainty measures, and assess
indexes' robustness. The paper presents three examples to illustrate the
usage of the pipeline framework: comparison of two different indexes
designed to monitor the spatio-temporal distribution of drought in
Queensland, Australia; the effect of dimension reduction choices on the
Global Gender Gap Index (GGGI) on countries' ranking; and how to
calculate bootstrap confidence intervals for the Standardized
Precipitation Index (SPI). The methods are supported by a new R package,
called \texttt{tidyindex}. Supplemental materials for the article are
available online.
\end{abstract}
\begin{keywords}
\def\sep{;\ }
indexes\sep data pipeline\sep software design\sep uncertainty\sep 
decision-making
\end{keywords}

\setstretch{2}
\section{Introduction}\label{introduction}

Indexes are commonly used to combine and summarize different sources of
information into a single number for monitoring, communicating, and
decision-making. They serve as critical tools across the natural and
social sciences. Examples include the Air Quality Index, El
Niño-Southern Oscillation Index, Consumer Price Index, QS University
Rankings, and the Human Development Index. In environmental science,
climate indexes are produced by major monitoring centers, like the
United States Drought Monitor and National Oceanic and Atmospheric
Administration, to facilitate agricultural planning and early detection
of natural disasters. In economics, indexes provide insight into market
trends through combining prices of a basket of goods and services. In
social sciences, indexes are used to monitor human development, gender
equity, or university quality.

The problem is that every index is developed in its own unique way, by
different researchers or organizations, and often indexes designed for
the same purpose cannot easily be compared. This echoes an issue raised
in Donoho (2017), that different data analysts might arrive at different
conclusions despite using the same data. This is especially pertinent to
index use which affects important decisions such as in natural disaster
prevention, economic interventions, resource allocation or human
development. It is primarily due to a lack of standards in data analysis
workflow. Donoho (2017) called for research on structuring a unified
workflow to address methodological variation across studies in data
science. Current practices also violate statistical principles, where
quantifying and understanding uncertainty are essential to deciding on a
best measure or metric, for example by incorporating bootstrap
confidence intervals (Efron 1979). There has been considerable research
in tidying up routine data analyses (Wickham 2014, 2011; Kuhn and Silge
2022; Wang, Cook, and Hyndman 2020; Zhang et al. to appear) that ``turn
ideas into software quickly and faithfully'', as envisioned by Chambers
(1998). Index development and use needs tidying.

To construct an index, experts typically start by defining a concept of
interest that requires measurement. This concept often lacks a direct
measurable attribute or can only be measured as a composite of various
processes, yet it holds social and public significance. To create an
index, once the underlying processes involved are identified, relevant
and available variables are then defined, collected, and combined using
statistical methods into an index that aims to measure the process of
interest. The construction process is often not straightforward, and
decisions need to be made, such as the selection of variables to be
included, which might depend on data availability and the statistical
definition of the index to be used, among others. For instance, the
indexes constructed from a linear combination of variables require a
decision on the weight assigned to each variable. Some indexes have a
spatial and/or temporal component, and variables can be aggregated to
different spatial resolutions and temporal scales, leading to various
indexes for different monitoring purposes. Hence, all these decisions
can result in different index values and have different practical
implications.

To be able to test different decision choices for an index,
systematically and statistically, the index needs to be broken down into
its fundamental building blocks to analyze the contribution and effect
of each component. We call this process the \emph{index pipeline}, which
are the steps of the data analysis pipeline for index construction. Such
a decomposition of index components provides the means to standardize
index construction via a pipeline and offers benefits for comparing
versions of indexes, calculating index uncertainty, and assessing index
robustness. It also provides clear recipes for the index definition,
facilitating reproducibility of results.

Here we detail the statistical and computational methods for developing
a data pipeline framework to construct and customize indexes using data.
The pipeline comprises various modules, including temporal and spatial
aggregation, variable transformation and combination, distribution
fitting, benchmark setting, and index communication. When combining
multivariate data into indexes, the pipeline enables the evaluation of
how any particular combination can affect the index. Uncertainty
calculation can also flow through the pipeline to provide an index with
confidence intervals. The pipeline also fits neatly into current tidy
data workflows and data visualisation.

The rest of the paper is structured as follows.
Section~\ref{sec-idx-dev} provides background about the development of
indexes. Section~\ref{sec-tidy} reviews the tidy framework in R and how
index construction can benefit from such a framework. The details of the
pipeline modules are presented in Section~\ref{sec-pipeline}.
Section~\ref{sec-software} explains the design of the \texttt{tidyindex}
package that implements the modules. Examples are given in
Section~\ref{sec-examples} to illustrate three use cases of the
pipeline.

\section{Background to index development}\label{sec-idx-dev}

There are many documents providing advice on how to construct indexes
for different fields, and review articles describing the range of
available indexes for specific purposes. The OECD handbook (OECD,
European Union, and Joint Research Centre - European Commission 2008)
provides a comprehensive guide for computing socio-economic composite
indexes, with detailed steps and recommendations. The drought index
handbook (Svoboda, Fuchs, et al. 2016) provides details of various
drought indexes and recommendations from the World Meteorology
Organization. Zargar et al. (2011), Hao and Singh (2015) and Alahacoon
and Edirisinghe (2022) are review papers describing the range of
possible drought indexes.

There is also some attention being given to the diagnosis of indexes,
and incorporation of uncertainty. Jones and Andrey (2007) investigates
the methodological choices made in the development of indexes for
assessing vulnerable neighborhoods. Saisana, Saltelli, and Tarantola
(2005) describes incorporating uncertainty estimates and conducting
sensitivity analysis on composite indexes. Tate (2012) and Tate (2013),
similarly, make a comparative assessment of social vulnerability indexes
based on uncertainty estimation and sensitivity analysis. Laimighofer
and Laaha (2022) studies five uncertainty sources (record length,
observation period, distribution choice, parameter estimation method,
and GOF-test) of drought indexes.

There are also a few R packages supporting index calculation. The
\texttt{SPEI} package (Beguería and Vicente-Serrano 2017) computes two
drought indexes. The \texttt{gpindex} package (Martin 2023) computes
price indexes, and the \texttt{fundiversity} package (Grenié and Gruson
2023) computes functional diversity indexes for ecological study. The
package \texttt{COINr} (Becker et al. 2022) is more ambitious, making a
start on following the broader guidelines in the OECD handbook to
construct, analyze, and visualize composite indexes.

From reviewing this literature, and in the process of developing methods
for making it easier to work with multivariate spatio-temporal data, it
seems possible to think about indexes in a more organised, cohesive and
standard manner. Actually, the area could benefit from a \emph{tidy}
approach.

\section{Tidy framework}\label{sec-tidy}

The tidy framework consists of two key components: tidy data and tidy
tools. The concept of tidy data (Wickham 2014) prescribes specific rules
for organizing data in an analysis, with observations as rows, variables
as columns, and types of observational units as tables. Tidy tools, on
the other hand, are concatenated in a sequence through which the tidy
data flows, creating a pipeline for data processing and modeling. These
pipelines are data-centric, meaning all the tidy tools or functions take
a tidy data object as input and return a processed tidy data object,
directly ready for the next operations to be applied. Also, the pipeline
approach corresponds to the modular programming practice, which breaks
down complex problems into smaller and more manageable pieces, as
opposed to a monolithic design, where all the steps are predetermined
and integrated into a single piece. The flexibility provided by the
modularity makes it easier to modify certain steps in the pipeline and
to maintain and extend the code base.

Examples of using a pipeline approach for data analysis can be traced
back to the interactive graphics literature, including A. Buja et al.
(1988); Sutherland et al. (2000); Wickham et al. (2009); Xie, Hofmann,
and Cheng (2014). Wickham et al. (2009) argue that whether made explicit
or not, a pipeline has to be presented in every graphics program, and
making them explicit is beneficial for understanding the implementation
and comparing between different graphic systems. While this comment is
made in the context of interactive graphics programs, it is also
applicable generally to any data analysis workflow. More recently, the
tidyverse suite (Wickham et al. 2019) takes the pipeline approach for
general-purpose data wrangling and has gained popularity within the R
community. The pipeline-style code can be directly read as a series of
operations applied successively on tidy data objects, offering a method
to document the data wrangling process with all the computational
details for reproducibility.

Since the success of tidyverse, more packages have been developed to
analyze data using the tidy framework for domain-specific applications,
a noticeable example of which is \texttt{tidymodels} for building
machine learning models (Kuhn and Silge 2022). To create a tidy workflow
tailored to a specific domain, developers first need to identify the
fundamental building blocks to create a workflow. These components are
then implemented as modules, which can be combined to form the pipeline.
For example, in supervised machine learning models, steps such as data
splitting, model training, and model evaluation are commonly used in
most workflows. In the \texttt{tidymodels}, these steps are
correspondingly implemented as packages \texttt{rsample},
\texttt{parsnip}, and \texttt{yardstick}, agnostic to the specific model
chosen. The uniform interface in tidymodels frees analysts from
recalling model-specific syntax for performing the same operation across
different models, increasing the efficiency to work with different
models simultaneously.

For constructing indexes, the pipeline approach adopts explicit and
standalone modules that can be assembled in different ways. Index
developers can choose the appropriate modules and arrange them
accordingly to generate the data pipeline that is needed for their
purpose. The pipeline approach provides many advantages:

\begin{itemize}
\tightlist
\item
  makes the computation more transparent, and thus more easily debugged,
  facilitating reproducibility.
\item
  allows for rapidly processing new data to check how different
  features, like outliers, might affect the index value.
\item
  provides the capacity to measure uncertainty by computing confidence
  intervals from multiple samples as generated by bootstrapping the
  original data.
\item
  enables systematic comparison of surrogate indexes designed to measure
  the same phenomenon.
\item
  it may even be possible to automate diagrammatic explanations and
  documentation of the index.
\end{itemize}

The adoption of this pipeline approach would provide uniformity to the
field of index development, research, and application to improve
comparability, reproducibility, and communication.

\section{Details of the index pipeline}\label{sec-pipeline}

In constructing various indexes, the primary aim is to transform the
data, often multivariate, into a univariate index. Spatial and temporal
considerations are also factored into the process when observational
units and time periods are not independent. However, despite the
variations in contextual information for indexes in different fields,
the underlying statistical methodology remains consistent across diverse
domains. Each index can be represented as a series of modular
statistical operations on the data. This allows us to decompose the
index construction process into a unified pipeline workflow with a
standardized set of data processing steps to be applied across different
indexes.

\begin{figure}

\centering{

\includegraphics[width=1\textwidth,height=0.9\textheight]{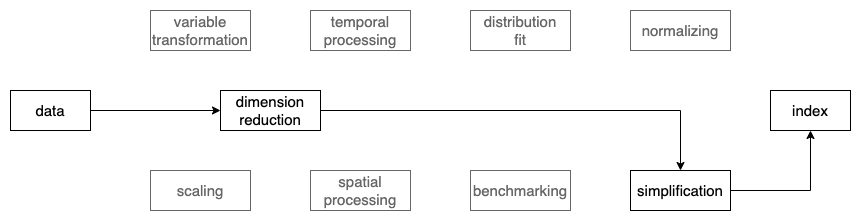}

}

\caption{\label{fig-pipeline-steps}Diagram of pipeline modules for index
construction. The highlighted path illustrates one possible construction
using the dimension reduction and simplification modules.}

\end{figure}

An overview of the pipeline is presented in
Figure~\ref{fig-pipeline-steps}, illustrating the nine available modules
designed to obtain the index from the data. These modules include
operations for temporal and spatial aggregation, variable transformation
and combination, distribution fitting, benchmark setting, and index
communication. Analysts have the flexibility to construct indexes by
connecting modules according to their preferences.

Now, we introduce the notation used for describing pipeline modules.
Consider a multivariate spatio-temporal process,
\[\mathbf{x}(s;t) = \{x_1(s;t), x_2(s;t), \cdots, x_p(s;t)\} \qquad s \in D_s \subseteq \mathbb{R}^m, t \in D_t \subseteq \mathbb{R}^n \]
where:

\begin{itemize}
\item
  \(x_j(s, t)\) represents a variable of interest for example
  precipitation, \(j = 1, \cdots, p\),
\item
  \(s\) represents the geographic locations in the space
  \(D_s \subseteq \mathbb{R}^m\). Examples of geographic locations
  include a collection of countries, longitude and latitude coordinates
  or regions of interest and,
\item
  \(t\) denotes the temporal order in \(D_t \subseteq \mathbb{R}^n\).
  For instance, time measurements could be recorded hourly, yearly,
  monthly, quarterly, or by season.
\end{itemize}

In what follows when geographic or temporal components of the
\(x_j(s,t)\) process are fixed we will be using suffix notation. For
example, \(x_{sj}(t)\) represents the data for a fixed location \(s\) as
a function of time \(t\), while \(x_{tj}(s)\) denotes the spatial
varying process for a fixed \(t\). An overview of the notation for
pipeline input, operation, and output is presented in
Table~\ref{tbl-notation}.

\begin{longtable}[]{@{}
  >{\raggedright\arraybackslash}p{(\columnwidth - 6\tabcolsep) * \real{0.3151}}
  >{\raggedright\arraybackslash}p{(\columnwidth - 6\tabcolsep) * \real{0.1233}}
  >{\raggedright\arraybackslash}p{(\columnwidth - 6\tabcolsep) * \real{0.2055}}
  >{\raggedright\arraybackslash}p{(\columnwidth - 6\tabcolsep) * \real{0.3562}}@{}}
\caption{Summary of the notation for input, operation, and output of
each pipeline module.}\label{tbl-notation}\tabularnewline
\toprule\noalign{}
\begin{minipage}[b]{\linewidth}\raggedright
Module
\end{minipage} & \begin{minipage}[b]{\linewidth}\raggedright
Input
\end{minipage} & \begin{minipage}[b]{\linewidth}\raggedright
Operation
\end{minipage} & \begin{minipage}[b]{\linewidth}\raggedright
Output
\end{minipage} \\
\midrule\noalign{}
\endfirsthead
\toprule\noalign{}
\begin{minipage}[b]{\linewidth}\raggedright
Module
\end{minipage} & \begin{minipage}[b]{\linewidth}\raggedright
Input
\end{minipage} & \begin{minipage}[b]{\linewidth}\raggedright
Operation
\end{minipage} & \begin{minipage}[b]{\linewidth}\raggedright
Output
\end{minipage} \\
\midrule\noalign{}
\endhead
\bottomrule\noalign{}
\endlastfoot
Temporal processing & \(x_{sj}(t)\) & \(f[x_{sj}(t)]\) &
\(x^{\text{Temp}}_{sj}(t^\prime) \quad t^\prime \in D_{t^\prime}\) \\
Spatial processing & \(x_{tj}(s)\) & \(g[x_{tj}(s)]\) &
\(x^{\text{Spat}}_{tj}(s^\prime) \quad s^\prime \in D_{s^\prime}\) \\
Variable transformation & \(x_{j}(s; t)\) & \(T[x_j(s;t)]\) &
\(x^{\text{Trans}}_j(s;t)\) \\
Scaling & \(x_j(s; t)\) & \([x_j(s;t) - \alpha]/\gamma\) &
\(x^{\text{Scale}}_j(s;t)\) \\
Dimension reduction & \(\mathbf{x}(s;t)\) & \(h[\mathbf{x}(s;t)]\) &
\(\mathbf{y}(s;t) \quad \mathbf{y} \subseteq \mathbb{R}^d, d < p\) \\
Distribution fit & \(x_j(s; t)\) & \(F[x_j(s;t)]\) &
\(P_j(s;t) \quad P(.) \in [0, 1]\) \\
Normalising & \(x_j(s; t)\) & \(\Phi^{-1}[x_j(s; t)]\) &
\(z_j(s; t)\) \\
Benchmarking & \(x_j(s; t)\) & \(u[x_j(s;t)]\) & \(b_j(s;t)\) \\
Simplification & \(x_j(s; t)\) & \(v[x_j(s;t)]\) &
\(A_j(s;t) \in \{a_1, a_2, \cdots, a_z\}\) \\
\end{longtable}

\subsection{Temporal processing}\label{temporal-processing}

The temporal processing module takes as input argument a single variable
\(x_{sj}(t)\) at location \(s\) as a function of time. In this step, the
original time series can be transformed or summarized into a new one via
time aggregation. The transformation is represented by the function
\(f\), \(x^{\text{Temp}}_{sj}(t^\prime) = f[x_{sj}(t)]\) where
\(t^\prime\) refers to the new temporal resolution after aggregation. An
example of temporal processing done in the computation of the
Standardized Precipitation Index (SPI) (McKee et al. 1993), consists of
summing the monthly precipitation series over a rolling time window of
size \(k\). That is also known as the time scale. For SPI, the choice of
the time scale \(k\) is used to control the accumulation period for the
water deficit, enabling the assessment of drought severity across
various types (meteorological, agricultural, and hydrological).

\subsection{Spatial processing}\label{spatial-processing}

The spatial processing module takes a single variable with a fixed
temporal dimension, \(x_{tj}(s)\), as input. This step transforms the
variable from the original spatial dimension \(s\) into the new
dimension \(s^\prime \in D_{s^\prime}\) through
\(x^{\text{Spat}}_{tj}(s^\prime) = g[x_{tj}(s)]\) via a function \(g\).
The change of spatial dimension allows for the alignment of variables
collected from different measurements, such as in-situ stations and
satellite imagery, or originating from different resolutions. This also
includes the aggregation of variables into different levels, such as
city, state, and country scales.

\subsection{Variable transformation}\label{variable-transformation}

Variable transformation takes the input of a single variable
\(x_j(s;t)\) and reshapes its distribution using the function \(T\) to
produce \(x^{\text{Trans}}_{j}(s;t)\). When a variable has a skewed
distribution, transformations such as log, square root, or cubic root
can adjust the distribution towards normality. For example, in the Human
Development Index (HDI), a logarithmic transformation is applied to the
variable Gross National Income per capita (GNI), to reduce its impact on
HDI, particularly for countries with high GNI values.

\begin{figure}

\centering{

\includegraphics{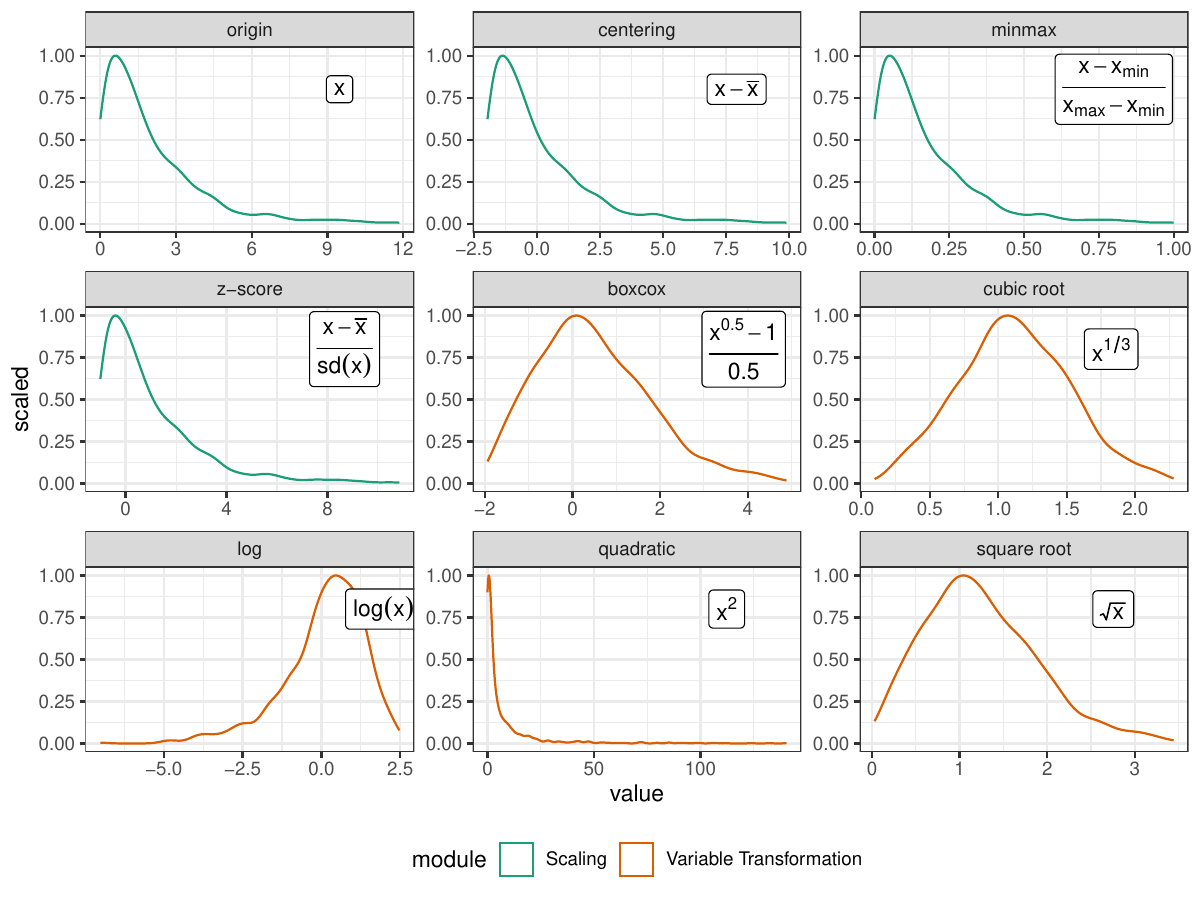}

}

\caption{\label{fig-scale-var-trans-compare}Comparison of the scaling
(green) and variable transformation (orange) modules. While both modules
change the variable range, scaling maintains the same distributional
shape, which is not the case with variable transformation.}

\end{figure}%

\subsection{Scaling}\label{scaling}

Unlike variable transformation, scaling maintains the distributional
shape of the variable. It includes techniques such as centering, z-score
standardization, and min-max standardization and can be expressed as
\([x_{j}(s;t) - \alpha]/\gamma\) where \(\alpha\) and \(\gamma\) are
constants. In the Human Development Index (HDI), the three dimensions
(health, education, and economy) are converted into the same scale (0-1)
using min-max standardization.

Although the scaling might be considered to be a transformation, we have
elected to make it a separate module because it is neater.
Figure~\ref{fig-scale-var-trans-compare} shows that scaling simply
changes the numbers in the data but not the shape of a variable, while
transformation will most likely change the shape, as it is usually
non-linear.

\subsection{Dimension reduction}\label{dimension-reduction}

Dimension reduction takes the multivariate information
\(\mathbf{x}(s;t)\), where \(\mathbf{x} \subseteq \mathbb{R}^p\), or a
subset of variables \(x_i(s;t)\) in \(\mathbf{x}(s;t)\), as the input.
It summarises the high-dimensional information into a lower-dimension
representation \(\mathbf{y}(s;t)\), where
\(\mathbf{y} \subseteq \mathbb{R}^d\) and \(d < p\), as the output. The
transformation can be based on domain-specific knowledge, originating
from theories describing the underlying physical processes, or guided by
statistical methods. For example, the Standardized
Precipitation-Evapotranspiration Index (SPEI) (Vicente-Serrano,
Beguería, and López-Moreno 2010) calculates the difference \(D\) between
precipitation (\(P\)) and potential evapotranspiration (\(\text{PET}\)),
using a water balance model (\(D = P - \text{PET}\)). This is the only
step that differs from the Standardized Precipitation Index (SPI), and
can be considered to be a dimension reduction using a particular linear
combination.

Linear combinations of variables are commonly used to reduce the
dimension in statistical methodology, and chosen using a method like
principal component analysis (PCA) (Hotelling 1933) or linear
discriminant analysis (Ronald A. Fisher 1936), preparing contrasts to
test particular elements in analysis of variance (Ronald Aylmer Fisher
1970), or hand-crafted by a content-area expert. Linear combinations
also form the basis for visualizing multivariate data, in methods such
as tours (Wickham et al. 2011). This dimension reduction method can
accommodate linear combinations as provided by any method, and hence is
linear by design. The transformation module provides variable-wise
non-linear transformation.

\subsection{Distribution fit}\label{distribution-fit}

Distribution fit applies the Cumulative Distribution Function (CDF)
\(F\) of a distribution on the variable \(x_j(s; t)\) to obtain the
probability values \(P_j(s;t) \in [0, 1]\). In SPEI, many distributions,
including log-logistic, Pearson III, lognormal, and general extreme
distribution, are candidates for the aggregated series. Different
fitting methods and different goodness of fit tests may be used to
compare the distribution choice on the index value. This could be
considered to be a variable transformation because it is usually
conducted separately for each variable. However, very occasionally a fit
is conducted on two or more variables simultaneously. For this reason,
and because it usually is applied later in the pipeline it is neater to
make this a separate module.

\subsection{Normalising}\label{normalising}

Normalizing applies the inverse normal CDF \(\Phi^{-1}\) on the input
data to obtain the normal density \(z_{j}(s;t)\). Normalizing can
sometimes be confused with the scaling or variable transformation
module, which does not involve using a normal distribution to transform
the variable. It is arguably whether normalizing and distribution fit
should be combined or separated into two modules. A decision has been
made to separate them into two modules given the different types of
output each module presents (probability values for distribution fit and
normal density values for normalizing).

\subsection{Benchmarking}\label{benchmarking}

Benchmark sets a value \(b_j(s,t)\) for comparing against the original
variable \(x_j(s;t)\). This benchmark can be a fixed value consistently
across space and time, perhaps extracted from expert knowledge or
determined by the data through the function \(u[x_j(s;t)]\). Once a
benchmark is set, observations can be highlighted for adjustments in
other modules or can serve as targets for monitoring and planning.

\subsection{Simplification}\label{simplification}

Simplification takes a continuous variable \(x_j(s;t)\) and categorises
it into a discrete set \(A_j(s;t) \in \{a_1, a_2, \cdots, a_z\}\)
through a piecewise constant function,

\begin{equation}
v[x_i(s;t)] = 
\begin{cases}
a_0, & C_1 \leq x^i(s; t) < C_0 \\
a_1, & C_2 \leq x^i(s; t) < C_1 \\
a_2, & C_3 \leq x^i(s; t) < C_2 \\
\cdots \\
a_z, & C_z \leq x^i(s; t)
\end{cases}
\end{equation}

This is typically used at the end of the index pipeline to simplify the
index to communicate to the public the severity of the concept of
interest measured by the index. An example of simplification is to map
the calculated SPI to four categories: mild, moderate, severe, and
extreme drought.

\section{Software design}\label{sec-software}

The R package \texttt{tidyindex} implements a proof-of-concept of the
index pipeline modules described in Section~\ref{sec-pipeline}. These
modules compute an index in a sequential manner, as shown below:

\begin{Shaded}
\begin{Highlighting}[]
\NormalTok{DATA }\SpecialCharTok{|\textgreater{}} \FunctionTok{module1}\NormalTok{(...) }\SpecialCharTok{|\textgreater{}} \FunctionTok{module2}\NormalTok{(...) }\SpecialCharTok{|\textgreater{}} \FunctionTok{module3}\NormalTok{(...) }\SpecialCharTok{|\textgreater{}}\NormalTok{ ...}
\end{Highlighting}
\end{Shaded}

Each module offers a variety of alternatives, each represented by a
distinct function. For example, within the
\texttt{dimension\_reduction()} module, three methods are available:
\texttt{aggregate\_linear()}, \texttt{aggregate\_geometrical()}, and
\texttt{manual\_input()} and they can be used as:

\begin{Shaded}
\begin{Highlighting}[]
\FunctionTok{dimension\_reduction}\NormalTok{(}\AttributeTok{V1 =} \FunctionTok{aggregate\_linear}\NormalTok{(...))}
\FunctionTok{dimension\_reduction}\NormalTok{(}\AttributeTok{V2 =} \FunctionTok{aggregate\_geometrical}\NormalTok{(...))}
\FunctionTok{dimension\_reduction}\NormalTok{(}\AttributeTok{V3 =} \FunctionTok{manual\_input}\NormalTok{(...))}
\end{Highlighting}
\end{Shaded}

Each method can be independently evaluated as a recipe, for example,

\begin{Shaded}
\begin{Highlighting}[]
\FunctionTok{manual\_input}\NormalTok{(}\SpecialCharTok{\textasciitilde{}}\NormalTok{x1 }\SpecialCharTok{+}\NormalTok{ x2)}
\end{Highlighting}
\end{Shaded}

takes a formula to combine the variables \texttt{x1} and \texttt{x2} and
return:

\begin{verbatim}
[1] "manual_input"
attr(,"formula")
[1] "x1 + x2"
attr(,"class")
[1] "dim_red"
\end{verbatim}

This recipe will then be evaluated in the pipeline module with data to
obtain numerical results. The package also offers wrapper functions that
combine multiple steps for specific indexes. For instance, the
\texttt{idx\_spi()} function bundles three steps (temporal aggregation,
distribution fit, and normalizing) into a single command, simplifying
the syntax for computation. Analysts are also encouraged to create
customized indexes from existing modules.

\begin{Shaded}
\begin{Highlighting}[]
\NormalTok{idx\_spi }\OtherTok{\textless{}{-}} \ControlFlowTok{function}\NormalTok{(...)\{}
\NormalTok{  DATA }\SpecialCharTok{|\textgreater{}} \FunctionTok{temporal\_aggregate}\NormalTok{(...) }\SpecialCharTok{|\textgreater{}} \FunctionTok{distribution\_fit}\NormalTok{(...)}\SpecialCharTok{|\textgreater{}} \FunctionTok{normalise}\NormalTok{(...)}
\NormalTok{\}}
\end{Highlighting}
\end{Shaded}

The tidyindex package is not intended to offer an exhaustive
implementation for all indexes across all domains. Instead, it provides
a realization of the pipeline framework proposed in the paper. When
adopting the pipeline approach to construct indexes, analysts may
consider developing software that can be readily deployed in the cloud
for production purposes.

\section{Examples}\label{sec-examples}

This section uses the example of drought and social indexes to show the
analysis made possible with the index pipeline. The drought index
example computes two indexes (SPI and SPEI) with various time scales and
distributions simultaneously using the pipeline framework to understand
the flood and drought events in Queensland. The second example focuses
on the dimension reduction step in the Global Gender Gap Index to
explore how the changes in linear combination weights affect the index
values and country rankings.

\subsection{Every distribution, every scale, every index all at
once}\label{sec-example1}

The state of Queensland in Australia frequently experiences natural
disaster events such as flood and drought, which can significantly
impact its agricultural industry. This example uses daily data from
Global Historical Climatology Network Daily (GHCND), aggregated into
monthly precipitation, to compute two drought indexes -- SPI and SPEI --
at various time scales and fitted distributions, for 29 stations in the
state of Queensland in Australia, spanning from January 1990 to April
2022. This example showcases the basic calculation of indexes with
different parameter specifications within the pipeline framework. The
dataset used in this example is available in the \texttt{tidyindex}
package as \texttt{queensland} and below we show the first few rows of
the data:

\begin{verbatim}
# A tibble: 5 x 9
  id                ym  prcp  tmax  tmin  tavg  long   lat name       
  <chr>          <mth> <dbl> <dbl> <dbl> <dbl> <dbl> <dbl> <chr>      
1 ASN00029038 1990 Jan  1682  34.3  24.7  29.5  142. -15.5 KOWANYAMA ~
2 ASN00029038 1990 Feb   416  35.2  23.2  29.2  142. -15.5 KOWANYAMA ~
3 ASN00029038 1990 Mar  2026  32.5  23.6  28.0  142. -15.5 KOWANYAMA ~
4 ASN00029038 1990 Apr   597  32.9  17.7  25.3  142. -15.5 KOWANYAMA ~
5 ASN00029038 1990 May   244  31.8  20.1  25.9  142. -15.5 KOWANYAMA ~
\end{verbatim}

\begin{figure}

\centering{

\includegraphics[width=1\textwidth,height=0.9\textheight]{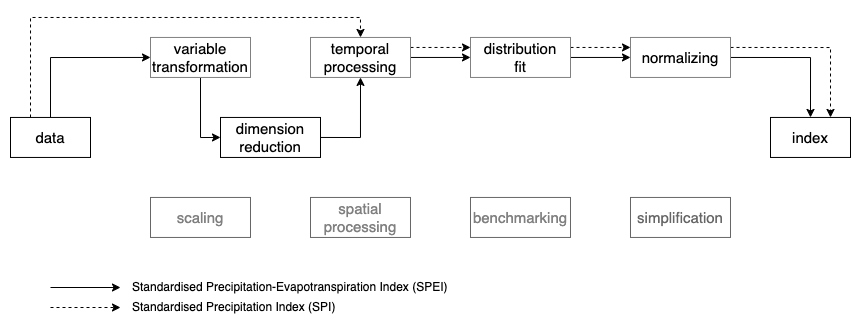}

}

\caption{\label{fig-spei}Index pipeline for two drought indexes: the
Standardized Precipitation Index (SPI) and the Standardized
Precipitation-Evapotranspiration Index (SPEI). Both indexes share
similar construction steps with SPEI having two additional steps
(variable transformation and dimension reduction) to convert temperature
into evapotranspiration and combine it with the precipitation series.}

\end{figure}%

Figure~\ref{fig-spei} illustrates the pipeline steps of the two indexes.
The two indexes are similar with the difference that SPEI involves two
additional steps -- variable transformation and dimension reduction --
prior to temporal processing. As introduced in
Section~\ref{sec-software}, wrapper functions are available for both
indexes as \texttt{idx\_spi()} and \texttt{idx\_spei()}, which allows
for the specification of different time scales and distributions for
fitting the aggregated series. In \texttt{tidyindex}, multiple indexes
can be calculated collectively using the function
\texttt{compute\_indexes()}. Both SPI and SPEI are calculated across
four time scales (6, 12, 24, and 36 months). The SPEI is fitted with two
distributions (log-logistic and general extreme value distribution) and
the gamma distribution is used for SPI:

\begin{Shaded}
\begin{Highlighting}[]
\NormalTok{.scale }\OtherTok{\textless{}{-}} \FunctionTok{c}\NormalTok{(}\DecValTok{6}\NormalTok{, }\DecValTok{12}\NormalTok{, }\DecValTok{24}\NormalTok{, }\DecValTok{36}\NormalTok{)}
\NormalTok{idx }\OtherTok{\textless{}{-}}\NormalTok{ queensland }\SpecialCharTok{\%\textgreater{}\%}
  \FunctionTok{mutate}\NormalTok{(}\AttributeTok{month =}\NormalTok{ lubridate}\SpecialCharTok{::}\FunctionTok{month}\NormalTok{(ym)) }\SpecialCharTok{|\textgreater{}}
  \FunctionTok{init}\NormalTok{(}\AttributeTok{id =}\NormalTok{ id, }\AttributeTok{time =}\NormalTok{ ym, }\AttributeTok{group =}\NormalTok{ month) }\SpecialCharTok{|\textgreater{}} 
  \FunctionTok{compute\_indexes}\NormalTok{(}
    \AttributeTok{spei =} \FunctionTok{idx\_spei}\NormalTok{(}
      \AttributeTok{.tavg =}\NormalTok{ tavg, }\AttributeTok{.lat =}\NormalTok{ lat, }
      \AttributeTok{.scale =}\NormalTok{ .scale, }\AttributeTok{.dist =} \FunctionTok{list}\NormalTok{(}\FunctionTok{dist\_gev}\NormalTok{(), }\FunctionTok{dist\_glo}\NormalTok{())),}
    \AttributeTok{spi =} \FunctionTok{idx\_spi}\NormalTok{(}\AttributeTok{.scale =}\NormalTok{ .scale)}
\NormalTok{  )}
\end{Highlighting}
\end{Shaded}

We use the \texttt{dplyr::glimpse()} function to inspect the
\texttt{idx} object created:

\begin{verbatim}
Rows: 128,576
Columns: 18
$ .idx   <chr> "spei", "spei", "spei", "spei", "spei", "spei", "spei~
$ .dist  <chr> "gev", "gev", "gev", "gev", "gev", "gev", "gev", "gev~
$ id     <chr> "ASN00029038", "ASN00029038", "ASN00029038", "ASN0002~
$ month  <dbl> 6, 7, 8, 9, 10, 11, 12, 12, 1, 1, 2, 2, 3, 3, 4, 4, 5~
$ ym     <mth> 1990 Jun, 1990 Jul, 1990 Aug, 1990 Sep, 1990 Oct, 199~
$ prcp   <dbl> 170, 102, 0, 0, 0, 278, 1869, 1869, 5088, 5088, 8484,~
$ tmax   <dbl> 29.65357, 31.20323, 31.32581, 32.80870, 36.80357, 36.~
$ tmin   <dbl> 16.25000, 17.15161, 13.11613, 16.25714, 21.49655, 24.~
$ tavg   <dbl> 22.95179, 24.17742, 22.22097, 24.53292, 29.15006, 30.~
$ long   <dbl> 141.7483, 141.7483, 141.7483, 141.7483, 141.7483, 141~
$ lat    <dbl> -15.4818, -15.4818, -15.4818, -15.4818, -15.4818, -15~
$ name   <chr> "KOWANYAMA AIRPORT", "KOWANYAMA AIRPORT", "KOWANYAMA ~
$ .pet   <dbl> 67.46933, 86.64868, 63.27450, 94.93572, 204.63793, 24~
$ .diff  <dbl> 102.53067, 15.35132, -63.27450, -94.93572, -204.63793~
$ .scale <chr> "6", "6", "6", "6", "6", "6", "6", "12", "6", "12", "~
$ .agg   <dbl> 4263.7863, 2819.8773, 2529.0243, 578.8843, -117.6571,~
$ .fit   <dbl> 0.02902164, 0.10512807, 0.57680687, 0.83297600, 0.818~
$ .index <dbl> -1.89537090, -1.25286143, 0.19373133, 0.96599235, 0.9~
\end{verbatim}

\begin{figure}

\centering{

\includegraphics{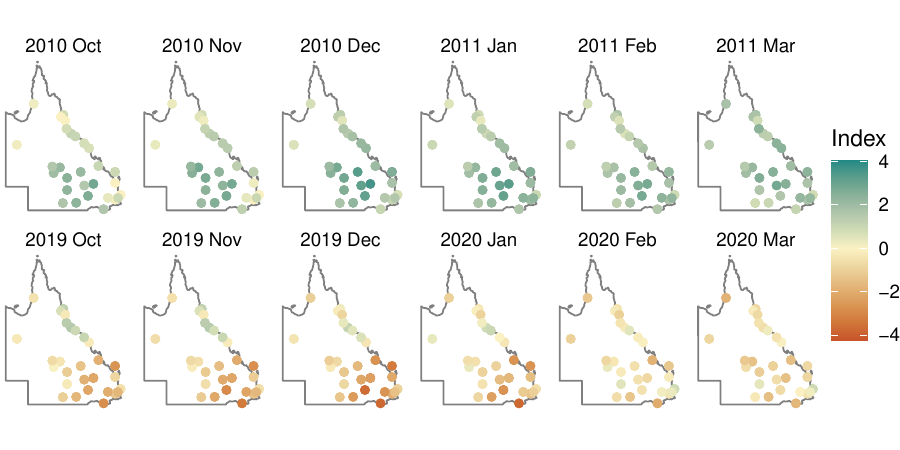}

}

\caption{\label{fig-compute-spatial}Spatial distribution of Standardized
Precipitation Index (SPI-12) in Queensland, Australia during two major
flood and drought events: 2010/11 and 2019/20. The map shows a
continuous wet period during the 2010/11 flood period and a mitigated
drought situation, after its worst in 2019 December and 2020 January,
likely due to the increased rainfall in February from the meteorological
record.}

\end{figure}%

\begin{figure}

\centering{

\includegraphics{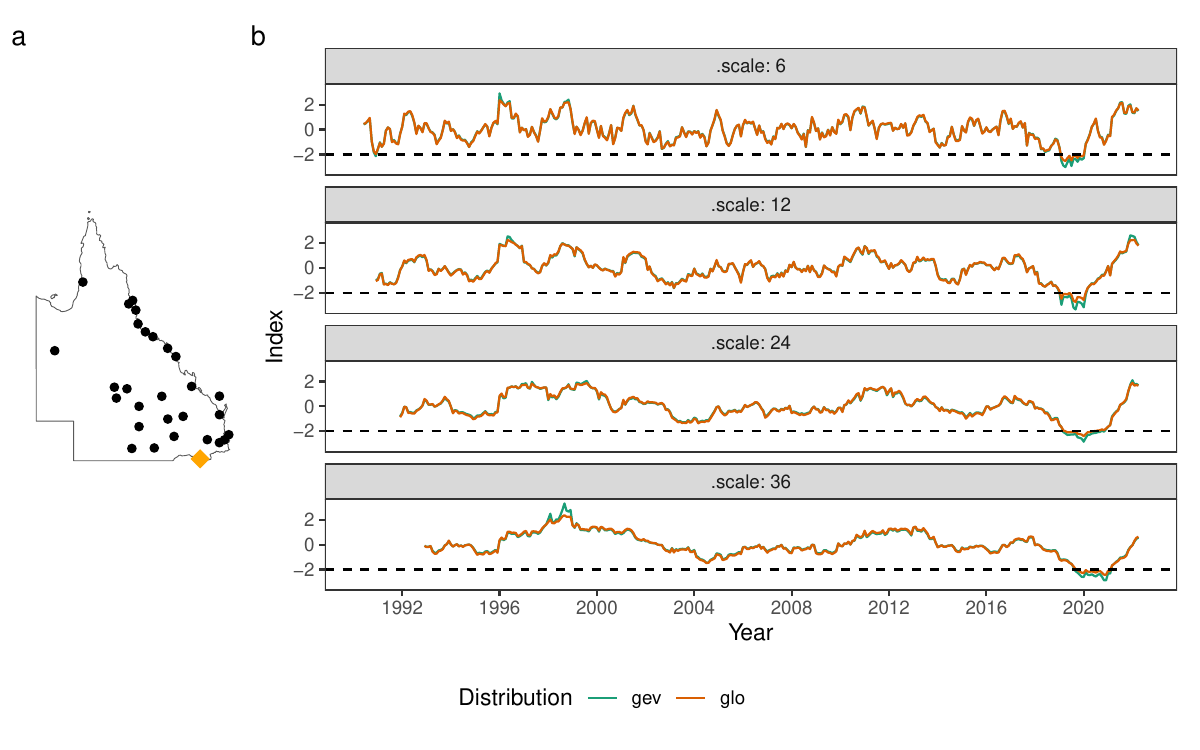}

}

\caption{\label{fig-compute-temporal}Time series plot of Standardized
Precipitation-Evapotranspiration Index (SPEI) at the Texas post office
station (highlighted by a diamond shape in panel a). The SPEI is
calculated at four time scales (6, 12, 24, and 36 months) and fitted
with two distributions (Log Logistic and GEV). The dashed line at -2
represents the class ``extreme drought'' by the SPEI. A larger time
scale gives a smoother index series, while also taking longer to recover
from an extreme situation as seen in the 2019/20 drought period. The
SPEI values from the two distributional fits mostly agree, while GEV can
result in more extreme values, i.e.~in 1998 and 2020.}

\end{figure}%

The output contains the original data, index values (\texttt{.index}),
parameters used (\texttt{.scale}, \texttt{.method}, and \texttt{.dist}),
and all the intermediate variables (\texttt{.pet}, \texttt{.agg}, and
\texttt{.fitted}). This data can be visualized to investigate the
spatio-temporal distribution of the drought or flood events, as well as
the response of index values to different time scales and distribution
parameters at specific single locations.
Figure~\ref{fig-compute-spatial} and Figure~\ref{fig-compute-temporal}
exemplify two possibilities. Figure~\ref{fig-compute-spatial} presents
the spatial distribution of SPI during two periods: October 2010 to
March 2011 for the 2010/11 Queensland flood and October 2019 to March
2020 for the 2019 Australia drought, which contributes to the notorious
2019/20 bushfire season. Figure~\ref{fig-compute-temporal} displays the
sensitivity of the SPEI series at the Texas post office to different
time scales and fitted distributions. Larger time scales produce a
smoother index across time, however, all time scales indicate an extreme
drought (corresponding to -2 in SPEI) in 2020, confirming the severity
of the drought across different time horizons. Moreover, the chosen
distribution has less influence on the index, with general extreme value
distribution tending to produce more extreme outcomes than log-logistic
distribution for the extreme events (index \textgreater{} 2 or
\textless-2).

\subsection{Does a puff of change in variable weights cause a tornado in
ranks?}\label{does-a-puff-of-change-in-variable-weights-cause-a-tornado-in-ranks}

\begin{figure}

\centering{

\includegraphics[width=1\textwidth,height=0.9\textheight]{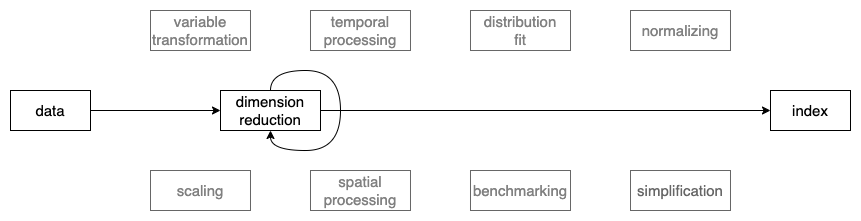}

}

\caption{\label{fig-pp-gggi}Index pipeline for the Global Gender Gap
Index (GGGI). The index is constructed as applying the module dimension
reduction twice on the data.}

\end{figure}

The Global Gender Gap Index (GGGI), published annually by the World
Economic Forum, measures gender parity by assessing relative gaps
between men and women in four key areas: Economic Participation and
Opportunity, Educational Attainment, Health and Survival, and Political
Empowerment (World Economic Forum 2023). The index, defined on 14
variables measuring female-to-male ratios, first aggregates these
variables into four dimensions (using the linear combination given by
\texttt{V-wgt} in Table~\ref{tbl-gggi-weights}). The weights are the
inverse of the standard deviation of each variable, scaled to sum to 1,
thus ensuring equal relative contribution of each variable to each of
the four new variables. These new variables are then combined through
another linear combination (\texttt{D-wgt} in
Table~\ref{tbl-gggi-weights}) to form the final index value.
Figure~\ref{fig-pp-gggi} illustrates that the pipeline is constructed by
applying the dimension reduction module twice on the data. The data for
GGGI does not needs to be transformed or scaled so these steps are not
included, but they might still need to be used for other similar
indexes.

\begingroup\fontsize{11}{13}\selectfont

\begin{longtable}[t]{>{\raggedright\arraybackslash}p{17.5em}>{\raggedleft\arraybackslash}p{4em}>{\raggedright\arraybackslash}p{3.5em}>{\raggedleft\arraybackslash}p{4em}>{\raggedleft\arraybackslash}p{3em}}

\toprule
\textbf{Variable} & \textbf{V-wgt} & \textbf{Dimension} & \textbf{D-wgt} & \textbf{weight}\\
\midrule
Labour force participation & 0.199 & Economy & 0.25 & 0.050\\
Wage equality for similar work & 0.310 &  &  & 0.078\\
Estimated earned income & 0.221 &  &  & 0.055\\
Legislators senior officials and managers & 0.149 &  &  & 0.037\\
Professional and technical workers & 0.121 &  &  & 0.030\\
\addlinespace
Literacy rate & 0.191 & Education & 0.25 & 0.048\\
Enrolment in primary education & 0.459 &  &  & 0.115\\
Enrolment in secondary education & 0.230 &  &  & 0.058\\
Enrolment in tertiary education & 0.121 &  &  & 0.030\\
Sex ratio at birth & 0.693 & Health & 0.25 & 0.173\\
\addlinespace
Healthy life expectancy & 0.307 &  &  & 0.077\\
Women in parliament & 0.310 & Politics & 0.25 & 0.078\\
Women in ministerial positions & 0.247 &  &  & 0.062\\
Years with female head of state & 0.443 &  &  & 0.111\\
\bottomrule

\caption{\label{tbl-gggi-weights}Weights for the two applications of
dimension reduction to compute the Global Gender Gap Index. V-wgt is
used to compute four new variables from the original 14. These are then
equally combined to get the final index value.}

\tabularnewline

\end{longtable}

\endgroup{}

The 2023 GGGI data is available from the Global Gender Gap Report 2023
in the country's economy profile and can be accessed in the
\texttt{tidyindex} package as \texttt{gggi} with
Table~\ref{tbl-gggi-weights} as \texttt{gggi\_weights}. The index can be
reproduced with:

\begin{Shaded}
\begin{Highlighting}[]
\NormalTok{gggi }\SpecialCharTok{\%\textgreater{}\%} 
  \FunctionTok{init}\NormalTok{(}\AttributeTok{id =}\NormalTok{ country) }\SpecialCharTok{\%\textgreater{}\%}
  \FunctionTok{add\_paras}\NormalTok{(gggi\_weights, }\AttributeTok{by =} \StringTok{"variable"}\NormalTok{) }\SpecialCharTok{\%\textgreater{}\%} 
  \FunctionTok{dimension\_reduction}\NormalTok{(}
    \AttributeTok{index\_new =} \FunctionTok{aggregate\_linear}\NormalTok{(}
      \SpecialCharTok{\textasciitilde{}}\NormalTok{labour\_force\_participation}\SpecialCharTok{:}\NormalTok{years\_with\_female\_head\_of\_state,}
      \AttributeTok{weight =}\NormalTok{ weight)) }
\end{Highlighting}
\end{Shaded}

After initializing the \texttt{gggi} object and attaching the
\texttt{gggi\_weights} as meta-data, a single linear combination within
the dimension reduction module is applied to the 14 variables (from
column \texttt{labour\_force\_participation} to
\texttt{years\_with\_female\_head\_of\_state}), using the weight
specified in the \texttt{wgt} column of the attached metadata. While
computing the index from the original 14 variables, it remains unclear
how the missing values are handled within the index, which impacts 68
out of the total 146 countries. However, after aggregating variables
into the four dimensions, where no missing values exist, the index is
reproducible for all the countries.

\begin{figure}

\centering{

\includegraphics{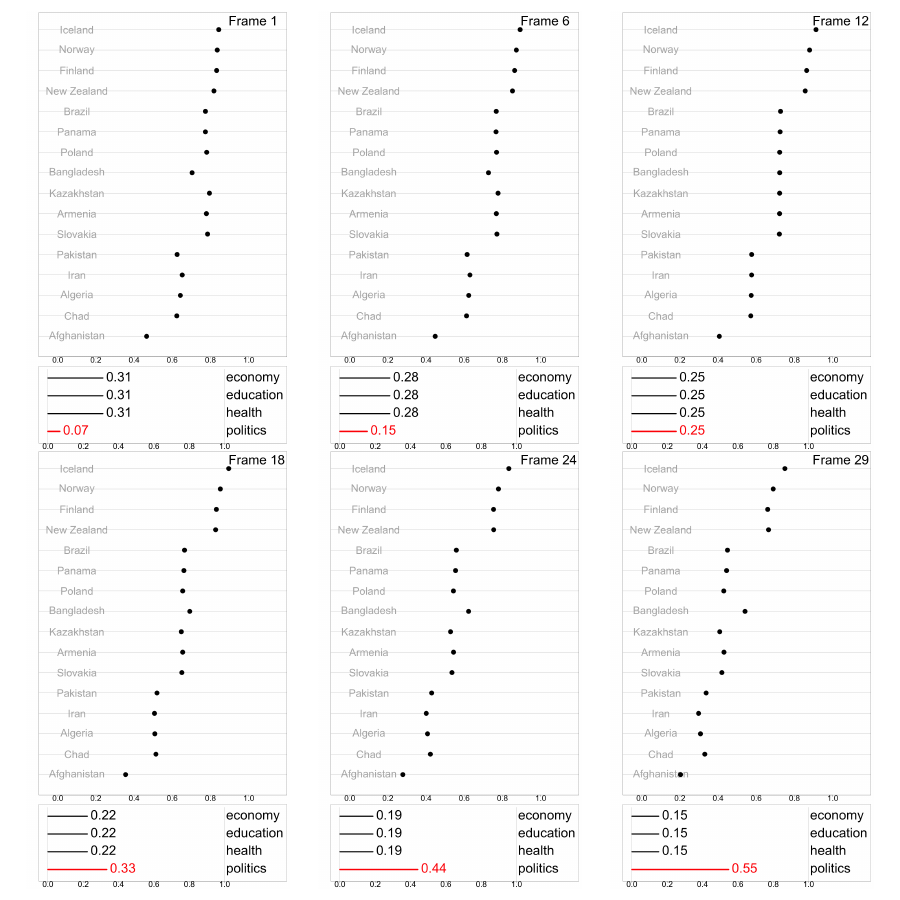}

}

\caption{\label{fig-idx-tour}Exploring the sensitivity of the GGGI, by
varying the politics component's contribution, for a subset of
countries. Each panel shows a dotplot of the index values, computed for
the linear combination represented by the segment plots below. Frame 12
shows the actual GGGI values, and countries are sorted from highest to
lowest on this. Frames 1 and 6 show the GGGI if the politics component
is reduced. Frames 18, 24, 29 show the GGGI when the politics component
is increased. The most notable feature is that Bangladesh's GGGI drops
substantially when politics is removed, indicating that this component
plays a large role in its relatively high value. Also, politics plays a
substantial role in the GGGI's for the top ranked countries, because
each of them drops, to the state of being similar to the middle ranked
countries when the politics component's contribution is reduced. The
animation can be viewed at https://vimeo.com/847874016.}

\end{figure}

Figure~\ref{fig-idx-tour} illustrates doing sensitivity analysis for
GGGI, for a subset of 16 countries. It presents 6 frames selected from
an animation where the weight on the politics dimension is gradually
increased, while other dimensions (economy, education, health) decrease
correspondingly. Frame 12 presents the original index where all the four
dimensions receive equal weight. The index values are sorted from
highest to lowest, with the Nordic countries (Iceland, Norway, and
Finland) and New Zealand leading the rankings. The index values are
between 0 and 1, and indicate proportional difference between men and
women, with a value of 0.8 indicating women are 80\% of the way to
equality of these measures. There is a gap in values between these
countries and the middle group (Brazil, Panama, Poland, Bangladesh,
Kazakhstan, Armenia, and Slovakia), and another big drop to the next
group (Pakistan, Iran, Algeria, and Chad). Afghanistan lags much further
behind.

To make a simple illustration of sensitivity analysis, we slightly vary
the weight for politics, between 0.07 and 0.52, while maintaining equal
weights among other dimensions. This can be viewed as an animation to
examine change in relative index values as a response to the changing
weights. This visualization technique, which presents a sequence of data
projections, is referred to as a ``tour'' and the specific kind of tour
used here to move between nearby projections is known as a ``radial
tour'' (see Andreas Buja et al. (2005), Wickham et al. (2011) and
Spyrison and Cook (2020) for more details).

Frames 1 and 6 show linear combinations where politics contributes less
than the original. It is interesting to note that the gap between the
Nordic countries and the middle countries dissipates, indicating that
this component was one reason for the relatively higher GGGI values of
these countries. Also interesting is the large drop in value for
Bangladesh. Frames 18, 24, 29 show linear combinations where politics
contributes more than the original. The most notable feature is that
Bangladesh retains its high index value whereas the other middle group
countries decline, indicating that the politics score is a major
component for Bangladesh's index value.

Ideally, an index should be robust against minor changes in its
construction components. This is not the case with GGGI, where small
changes to one component lead to fairly large change in the index. The
modular pipeline framework for computing the index makes it easy to
conduct this type of sensitivity analysis, where one or more components
are perturbed and the index recalculated. One aspect of the GGGI not
well-described in the Global Gender Gap Report is the handling of
missing values that are present in the initial variables for many
countries, something that is common for this type of data. This could
also be made more transparent with the dimension reduction module, by
specifying an imputation method or providing warnings about missing
values.

\subsection{Decoding uncertainty through the wisdom of the
crowd}\label{decoding-uncertainty-through-the-wisdom-of-the-crowd}

Errors in measurement, variability and sampling error, may arise at
various stages of the pipeline calculation, including from different
parameterization choices, as illustrated from
Section~\ref{sec-example1}, or from the statistical summarization
procedures applied in the pipeline. Although it may not be possible to
perfectly measure these errors, it is important that they are recognised
and estimated for an index, so that it is possible to compute confidence
intervals. In this example, the Texas post office station highlighted in
Figure~\ref{fig-compute-temporal} is used to illustrate one possibility
to compute a confidence interval for SPI. Bootstrapping is used to
account for the sampling uncertainty in the distribution fit step of the
index pipeline and to assess its impact on the SPI series.

In SPI, the distribution fit step fits the gamma distribution to the
aggregated precipitation series separately for each month. This results
in 32 or 33 points, from January 1990 to April 2022, for estimating each
set of distribution parameters. To account for this sampling uncertainty
with these samples, bootstrapping is used to generate replicates of the
aggregated series. In the \texttt{tidyindex} package, this bootstrap
sampling is activated when the argument \texttt{.n\_boot} is set to a
value other than the default of 1. In the following code, the
Standardized Precipitation Index (SPI) is calculated using a time scale
of 24. The bootstrap procedure samples the aggregated precipitation
(\texttt{.agg}) for 100 iterations (\texttt{.n\_boot\ =\ 100}) and then
fits the gamma distribution. The resulting gamma probabilities are then
transformed into normal densities in the normalizing step with
\texttt{normalise()}.

\begin{Shaded}
\begin{Highlighting}[]
\NormalTok{DATA }\SpecialCharTok{|\textgreater{}} 
  \FunctionTok{temporal\_aggregate}\NormalTok{(}\AttributeTok{.agg =} \FunctionTok{temporal\_rolling\_window}\NormalTok{(prcp, }\AttributeTok{scale =} \DecValTok{24}\NormalTok{)) }\SpecialCharTok{|\textgreater{}} 
  \FunctionTok{distribution\_fit}\NormalTok{(}\AttributeTok{.fit =} \FunctionTok{dist\_gamma}\NormalTok{(}\AttributeTok{var =} \StringTok{".agg"}\NormalTok{, }\AttributeTok{method =} \StringTok{"lmoms"}\NormalTok{,}
                                     \AttributeTok{.n\_boot =} \DecValTok{100}\NormalTok{)) }\SpecialCharTok{|\textgreater{}}
  \FunctionTok{normalise}\NormalTok{(}\AttributeTok{.index =} \FunctionTok{norm\_quantile}\NormalTok{(.fit))}
\end{Highlighting}
\end{Shaded}

\begin{figure}

\centering{

\includegraphics{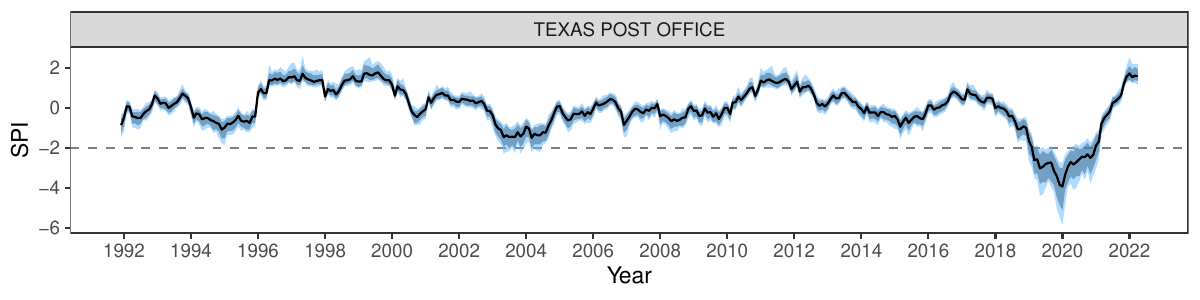}

}

\caption{\label{fig-conf-interval}80\% and 95\% confidence intervals of
the Standardized Precipitation Index (SPI-24) for the Texas post office
station, in Queensland, Australia. A bootstrap sample of 100 is taken
from the aggregated precipitation series to estimate gamma parameters
and to calculate the index. The dashed line at SPI = -2 represents an
extreme drought as defined by the SPI. Most parts of the confidence
intervals from 2019 to 2020 sit below the extreme drought line and are
relatively wide compared to other time periods. This suggests that while
it is certain that the Texas post office is suffering from a drastic
drought, there is considerable uncertainty in quantifying its severity,
given the extremity of the event.}

\end{figure}

The confidence interval can then be calculated using the quantile method
from the bootstrap samples. Figure~\ref{fig-conf-interval} presents the
80\% and 95\% confidence intervals for the Texas post office station, in
Queensland, Australia. From the start of 2019 to 2020, the majority of
the confidence intervals lie below the extreme drought line (SPI = -2),
suggesting a high level of certainty that the Texas post office is
suffering from a drastic drought. Also close to the extreme drought line
is the period 2003-2004, which corresponds to the millennium drought.
These relatively wide confidence intervals, as well as during the
excessive precipitation events in 1996-1998 and 1999-2000, suggest a
high variation of the gamma parameters estimated from the bootstrap
samples and its difficulty to accurately quantify the drought and flood
severity in extreme events.

\section{Conclusion}\label{conclusion}

The paper introduces a tidy data pipeline for constructing and analyzing
indexes. It has nine modules including temporal and spatial aggregation,
variable transformation and combination, distribution fitting, benchmark
setting, and index communication. This addresses statistical principles
absent from current index definitions: uncertainty quantification and
sensitivity. The tidyindex framework should encourage better statistical
practice wherever indexes are used as critical tools in natural and
social sciences.

Several examples are shown illustrating usage. For the drought indexes
(SPI and SPEI) we showed how multiple indexes can be computed with a
range of parameter choices, and compared across space and time. We
showed how bootstrap confidence intervals can be readily computed and
plotted to assess uncertainty about the index values, and how it may be
used to make better decisions on drought declarations. The Global Gender
Gap Index (GGGI) was used to illustrate how choices in dimension
reduction can radically affect index values and country rankings. This
also illustrates how the pipeline feeds nicely into advanced interactive
graphics.

There are many potential directions for the development of the work.
Computationally, the tidyindex framework could be extended to support
other data formats, like NetCDF for climate indexes. Conceptually,
extending the examples to re-express additional common-practice indexes
in the pipeline structure will help broader adoption, and further test
that the framework can indeed accommodate any and all possible indexes.

\section{Acknowledgement}\label{acknowledgement}

The steps of this pipeline are available in the R package,
\texttt{tidyindex}, available on CRAN. The source code for reproducing
the work reported in this paper are in Supplementary Materials and can
be found at: \url{https://github.com/huizezhang-sherry/paper-tidyindex}.

This work is funded by a Commonwealth Scientific and Industrial Research
Organisation (CSIRO) Data61 Scholarship. Nicolas Langrené acknowledges
the partial support of the Guangdong Provincial Key Laboratory IRADS
(2022B1212010006, R0400001-22) and the UIC Start-up Research Fund
UICR0700041-22. The article is created using Quarto (Allaire et al.
2022) in R (R Core Team 2021).

\section{Supplementary Materials}\label{supplementary-materials}

The supplementary materials include the full scripts for the three
examples in Section 6 (\texttt{/scripts} folder), a saved data for
Example 6.1 (\texttt{/data} folder), and a \texttt{README.md} file
containing the install instructions for running the scripts.

\section*{References}\label{references}
\addcontentsline{toc}{section}{References}

\phantomsection\label{refs}

\end{document}